\documentclass[12pt]{article}

\newsavebox{\ns}
\newsavebox{\dbrane}
\newsavebox{\dbshort}

\usepackage{epsfig}
\usepackage{amssymb}

\def\be{\begin{eqnarray}}
\def\ee{\end{eqnarray}}

\newcommand{\nn}{\nonumber}
\newcommand\para{\paragraph{}}
\newcommand{\ft}[2]{{\textstyle\frac{#1}{#2}}}
\newcommand{\eqn}[1]{(\ref{#1})}

\def\Dslash{\,\,{\raise.15ex\hbox{/}\mkern-12mu D}}
\def\dslash{\,\,{\raise.15ex\hbox{/}\mkern-12mu d}}
\def\Dbarslash{\,\,{\raise.15ex\hbox{/}\mkern-12mu {\bar D}}}
\def\delslash{\,\,{\raise.15ex\hbox{/}\mkern-9mu \partial}}
\def\delbarslash{\,\,{\raise.15ex\hbox{/}\mkern-9mu {\bar\partial}}}
\def\pslash{\,\,{\raise.15ex\hbox{/}\mkern-9mu p}}
\def\calDslash{\,\,{\raise.15ex\hbox{/}\mkern-12mu {\cal D}}}

\begin{document}
\pagestyle{plain}
\setcounter{page}{1}
\newcounter{bean}
\baselineskip16pt

\begin{titlepage}

\begin{center}
\today
\hfill hep-th/0212235\\
\hfill MIT-CTP-3334 \\

\vskip 1.5 cm
{\large \bf Comments on Condensates in} \\  
{\large \bf Non-Supersymmetric Orbifold Field Theories} \\ 
\vskip 1 cm 
{David Tong}\\
\vskip 1cm
{\sl Center for Theoretical Physics, 
Massachusetts Institute of Technology, \\ Cambridge, MA 02139, U.S.A.\\
{\tt dtong@mit.edu}}

\end{center}

\vskip 0.5 cm
\begin{abstract}
Non-supersymmetric orbifolds of ${\cal N}=1$ super Yang-Mills theories are 
conjectured to inherit properties from their supersymmetric parent. 
We examine this conjecture by compactifying the $Z_2$ orbifold theories on a 
spatial circle of radius $R$. We point out that when the orbifold theory lies 
in a specific weakly coupled vacuum, fractional instantons do give rise 
to the conjectured condensate of bi-fundamental fermions. However, we show 
that quantum effects render this vacuum unstable through the generation of 
twisted operators. In the true vacuum state, no fermion condensate forms. 
Thus, in contrast to super Yang-Mills, the compactified orbifold theory 
undergoes a chiral phase transition as $R$ is varied. 

\end{abstract}

\end{titlepage}

\subsubsection*{Introduction}

The holomorphic properties of supersymmetric gauge theories allow us 
to calculate certain quantities exactly, even in strongly 
coupled regimes. However, to make progress in more realistic, 
non-supersymmetric theories, we must learn to abandon our holomorphic 
comfort blanket. 

The  ``orbifold field theory'' is an interesting attempt in this direction 
which grew out of considerations in string theory. This method 
truncates a parent theory to a subset of fields which are left invariant under 
a discrete group action $G$. By suitably embedding $G$ in the parent 
R-symmetry group, non-supersymmetric daughter theories may be constructed 
from supersymmetric parents. 
Nevertheless, the daughter and parent theories enjoy the same planar graph expansion  
\cite{original,orig2,orig3,orig4}. It has been conjectured  by Strassler that this 
large N correspondence continues to hold for non-perturbative effects \cite{matt}. 
More precisely, the conjecture states that {\it if} the parent and orbifold theories 
share a common vacuum {\it then} the Green's functions for their shared operators 
will coincide. The 
importance of this conjecture lies in the fact that it maps the exactly 
computable correlation functions of a supersymmetric 
parent theory into its non-supersymmetric daughter. 

For the purpose of this short note, we take the parent theory to be ${\cal N}=1$ 
super Yang-Mills, with gauge group $U(2N)$. We choose to orbifold by 
a $G={\bf Z}_2$ action, which is embedded both within the gauge group and the 
non-anomalous ${\bf Z}_{4N}$ R-symmetry. The resulting daughter is \cite{ms},
\begin{center} $U(N)_1\times U(N)_2$ with a single Dirac fermion 
$\Psi=\left(\begin{array}{c}\lambda \\ \bar{\psi}\end{array}\right)$ 
transforming as $({\bf N},\bar{\bf N})$. 
\end{center}
For later convenience, we have decomposed the Dirac fermion into its 
Weyl constituents, $\lambda$ and $\psi$. Each gauge group has the same 
coupling constant $g^2$ and, at one-loop, the dynamically generated scale 
$\Lambda^3=\mu_0^3e^{-8\pi^2/g^2N}$ is chosen to coincide with that of the 
parent theory. The theta angles of the two groups are also set equal. 

The diagonal $U(1)$ of the gauge group decouples, while the remaining $U(1)$ 
acts as a gauged baryon number current. We shall denote it as $U(1)_B$. 
The classical theory enjoys a further global, chiral  
$U(1)_A$ symmetry, acting as $\Psi\rightarrow\exp(i\gamma_5\alpha)\Psi$. 
The anomaly ensures that only a ${\bf Z}_{N}$ subgroup survives quantisation. 
Finally, the theory also retains memory of its orbifold birth through a $G={\bf Z}_2$ global symmetry. 
Often referred to as the ``quantum symmetry'', it acts by exchanging the two gauge groups, 
while mapping $\lambda\leftrightarrow\psi$. Operators that carry charge under $G$ are known as 
``twisted operators'' and descend from non-gauge-invariant operators in the parent theory. 

For $N=3$, this theory is simply the chiral limit of QCD with three massless 
quarks, and the vector flavour group gauged. In analogy with QCD, it 
is expected that the chiral ${\bf Z}_N$ symmetry is spontaneously broken by 
a condensate of bi-fundamental fermions, resulting in $N$ vacuum states of the 
theory \cite{matt}. In the parent super Yang-Mills theory, the corresponding 
gluino condensate is exactly calculable and, assuming that the orbifold conjecture 
holds true, presents us with an exact prediction for the condensate in the orbifold 
theory \cite{matt,gs,dnv}
\be
\langle\lambda\psi\rangle=32\pi^2 \Lambda^3e^{i(\theta+2\pi k)/N}
\label{prediction}\ee
where $k=1,\ldots,N$ labels the $N$ degenerate vacua of the theory. 

It was 
pointed out in \cite{dnv} that, rather than discarding our holomorphic comfort blanket, 
we have instead wrapped a non-supersymmetric theory in it. This comment is motivated 
by the observation that the conjectured fermion condensate is holomorphic in the complexified 
coupling constant $\tau=\theta/2\pi+4\pi i/g^2$. This dependence is fixed by the 
one-loop RG flow, together with the anomaly. 

The original motivation of this paper was 
to test the prediction \eqn{prediction}. A direct 
strong coupling instanton calculation in four dimensions gives a non-zero 
result only for $\langle (\lambda\psi)^N\rangle$. This is a familiar story from 
super Yang-Mills theories where a useful strategy to overcome the problem 
is to compactify the theory on a spatial circle of radius $R$, endowing the fermions 
with periodic boundary 
conditions. Introducing a Wilson line has the dual advantage of 
making the theory weakly coupled, and introducing the relevant fractional 
instanton configurations, allowing for a controlled semi-classical 
calculation of the condensate 
\cite{swandur}. It is known that, as $R$ is varied, super 
Yang-Mills does {\em not} undergo a chiral phase transition and, in fact, the 
gluino condensate is independent of $R$. This is to 
be contrasted with thermal compactifications of this theory, in which the 
fermions have anti-periodic boundary conditions and a chiral phase transition 
does occur.

Since the perturbative orbifold conjecture is a property of planar graphs, one may 
expect it to continue to hold on ${\bf R}^{1,2}\times{\bf S}^1$. Moreover, the 
possibility of working at weak-coupling after compactification makes this a tempting 
path to follow, and was previously advocated in \cite{gs}.  Therefore, in this short 
note we shall consider the periodic compactification of the non-supersymmetric 
orbifold theory. In the next section we discuss perturbative results, turning 
to non-perturbative effects in the following section. We end with a short summary.

\subsubsection*{The Effective Potential and Vacuum Physics}

One advantage of compactifying, say, the $x^1$ direction on a circle of 
radius $R$ is that one may introduce Wilson lines to break the gauge group to 
the maximal torus and thus, for $R\ll 1/\Lambda$, to a weakly coupled theory. 
We work with the Lie algebra valued object,
\be
\int_0^{2\pi R} A^i_1\ dx^1 = {\rm diag}\,(v_1^i,\ldots,v_N^i)\ \ \ \ \ \ \ \ \ \ 
i=1,2
\label{poly}\ee
Large gauge transformations imply the periodicity of the eigenvalues, and we may take 
each to lie in the range $v_a^i\in[0,2\pi)$. The matter content of the theory is 
invariant under the 
diagonal $Z_N$ center of the two $SU(N)$ groups, allowing also for a twisted 
large gauge transformation. However, this is swamped by the overall $U(1)$ which 
we choose to keep.  The classical moduli space of vacua is therefore given by 
$(T^N/S_N)^2$, where $S_N$ is the Weyl group of $U(N)$. We may choose to fix the 
Weyl symmetry by insisting $v_a\leq v_{a+1}$. 
It will prove useful to depict the different classical vacua graphically as two 
distinct sets of eigenvalues distributed around a circle as shown in Figure 1.  
\begin{figure}
\begin{center}
\epsfxsize=4.5in\leavevmode\epsfbox{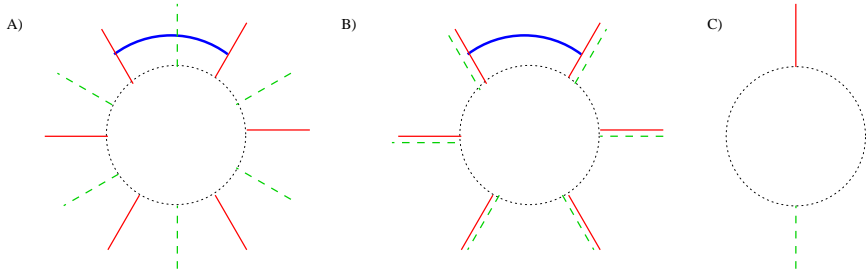}
\end{center}
\caption{\small{Three classical vacua of the theory. The solid (red) lines depict 
the ${\bf S}^1$ valued eigenvalues of the Wilson line for the $U(N)_1$ gauge 
group, while the dotted (green) lines depict the eigenvalues for $U(N)_2$. 
Calorons are shown as (blue) arcs connecting the eigenvalues of a 
given gauge group. This graphical representation finds life in the T-dual brane picture.}}
\end{figure}

Firstly, let us recall what becomes of the vacuum moduli space in the supersymmetric 
parent theory. One finds that the classical vacuum moduli space survives at the 
perturbative level. However, fractional instantons, of the type considered 
in the following section, act as a repulsive force, pushing apart the 
eigenvalues of the Wilson line \cite{vafa,swandur}. In this manner, the theory is 
driven to weak coupling by compactification. 

What does this supersymmetric vacuum descend to in the orbifold theory? In fact this 
point is already a little unclear. On graphical grounds, one may expect that the 
corresponding configuration consists of maximally separated, interlaced eigenvalues 
as depicted in Figure 1A
\be
v^1_a=v^2_a-\pi/N=2\pi a/N\ \ \ \ \ \ \ \ \ \ \ \ a=1,\ldots,N
\label{myvac}
\ee
However, in this vacuum the twisted operator $\int A_1-A_2$ has a non-vanishing 
expectation value.  To ensure that this operator vanishes, we should instead restrict 
to vacua of the daughter theory with $v_a^1=v_a^2$. If we insist that these eigenvalues 
are equally spaced, then we have the vacuum depicted in Figure 1B
\be
v_a^1=v_a^2=2\pi a/N\ \ \ \ \ \ \ \ \ \ \ \ a=1,\ldots,N
\label{myvac2}
\ee
However, as we shall now see, the discussion of which vacuum to pick is somewhat 
moot since the daughter theory will happily pick one for us. 
Given the matching of planar graphs \cite{original,orig2,orig3,orig4}, 
one may think that the perturbative stability of any vacuum state is similar 
to super Yang-Mills, at least at large $N$. This can be checked by computing 
the  one-loop contribution to the potential,
\be
{\cal F}=-\frac{1}{2\pi RV}\log 
\left(\frac{\det^2 (-D_f^2)}{\det (-D_1^2)\det (-D^2_2)}\right)
\label{f}\ee
where $V$ is the volume of the three-dimensional uncompactified space. The 
factors in the denominators come from integrating over the ghosts and 
gauge fields, with $D_i$ acting in the adjoint representation 
of the $U(N)_i$ gauge group. The numerator arises from integrating out 
the fermions, with $D_f$ in the bi-fundamental representation. 

To compute the determinants, we borrow commonplace techniques 
from thermal field theory, remembering that our fermions are endowed 
with periodic boundary conditions. The relevant computation may 
be found in Appendix D of \cite{gpy}, where the basic function 
$\Delta=\log\det(-(\partial_\mu+(v/2\pi R)\delta_{\mu 4})^2)$  
is calculated on ${\bf R}^3\times{\bf S}^1$ 
\be
\Delta&\equiv& \frac{V}{(2\pi)^3}
\sum_{n=-\infty}^\infty\int d^3k\ \log\left(k^2
+\frac{(2\pi n-v)^2}{(2\pi R)^2}\right)  \nn\\
&=& \Delta_0+\frac{2V}{(2\pi)^3}{\rm Re}\int_{-\infty+i\epsilon}^{+\infty+i\epsilon} 
dk_0\int d^3k\frac{\log(k_0^2+
{\bf k}\cdot{\bf k})}{\exp(-ik_0/2\pi R-iv)-1} \nn\\
&=& \Delta_0+\frac{2V}{(2\pi)^3}\int d^3k\,{\rm Re}\log\left(1-e^{-k/2\pi R+iv}\right) \nn\\
&=& 
\Delta_0-\frac{2V}{\pi^2}\frac{1}{(2\pi R)^3}\sum_{j=1}^\infty\frac{\cos(v j)}{j^4} 
\nn\ee
Here $\Delta_0$ is the divergent determinant evaluated on ${\bf R}^4$. It is 
independent of both $R$ and $v$, and is dealt with using a regularisation compatible with 
the supersymmetry of the parent theory, for example Pauli-Villars. If we take 
$v$ to lie in the range $v\in[0,2\pi)$, then the final sum may be explicitly performed, 
\be       
\Delta-\Delta_0 = -\frac{2V}{4\pi^5R^3}\left(
\frac{\pi^4}{90}-\frac{v^2}{48}(v-2\pi)^2\right)\ \ \ \ \ \ {\rm for\ }
v\in [0,2\pi)
\nn\ee
Using this result to calculate ${\cal F}$, we find that the $\Delta_0$ terms and 
the $\pi^4/90$ terms cancel between bosons and fermions. This is a manifestation of the 
parent supersymmetry. A finite contributions to the potential remains,
\be
{\cal F}&=&-\frac{1}{24\pi^2(2\pi R)^4}\sum_{a,b=1}^N
\left[2[v^1_a-v^2_b]^2([v^1_a-v^2_b]-2\pi)^2-[v^1_a-v^1_b]^2([v_a^1-v_b^1]-2\pi)^2
\right.\nn\\ &&\left.\hspace{1.5in}-[v^2_a-v^2_b]^2([v_a^2-v_b^2]-2\pi)^2\right]
\label{ff}\ee
where the square brackets are there to remind us that all periodically valued 
quantities live in the range $0<[v_a^i-v_b^j]\leq 2\pi$. The physics behind this 
potential is clear: the first term arises from the fermions which contribute a 
negative mass to the Wilson line and push the eigenvalues apart; the next two 
terms come from the gluons which contribute a positive mass, attracting the 
eigenvalues. In the supersymmetric theory, these two contributions cancel and 
no potential is generated at one-loop. Here however the bosons and fermions carry 
different gauge quantum numbers leading to the form \eqn{ff} for the potential. 
Similar potentials were found to lift flat directions in certain four-dimensional 
orbifold field theories with classical moduli spaces \cite{tk,as}. In that case, the 
potential arose from UV divergent effects through a Coleman-Weinberg mechanism. 
The potential \eqn{ff}, like those of \cite{tk,as}, involves a double 
trace term for twisted operators, and is not suppressed at large $N$. 
A simple way to understand 
that, despite planar equivalence, the orbifold theory doesn't obey its supersymmetric 
parent is to observe that the boson and fermion mass matrices differ in a generic 
vacuum.  In general, any orbifold theory with flat directions for twisted fields will 
suffer from a similar problem.

We may now judge the fate of the two vacua \eqn{myvac} and \eqn{myvac2}: both have  
vanishing vacuum energy, ${\cal F}=0$. Indeed, any vacuum in which the 
twisted operators vanish with $v_1^a=v_2^a$ has vanishing vacuum energy. 
Once again, this reflects the supersymmetric properties of the theories' ancestors. 
However, these supersymmetric configurations are not the vacua of the theory. 
This honour falls to the Wilson line depicted in Figure 1C,
\be
v^1_a=0\ {\rm and}\ v^2_a=\pi \ \ \ \ \ \ \ \ \ \ \ {\rm for}\ a=1,\ldots,N
\label{truevac}\ee
which boasts a vacuum energy of ${\cal F}=-N^2/3.2^6\pi^2 R^4$. 
\para
So what is the infra-red physics in the true vacuum? The full non-abelian gauge 
symmetry is restored, and the Wilson line \eqn{truevac} may be thought of as lying 
only in $U(1)_B$. This dynamically generated Wilson line gives the fermions 
a Kaluza-Klein mass $m=1/2R$. To see the effect of this mass, consider the 
decomposition of the fermions in Fourier modes along the compact circle  
$\Psi(x_1,x)=\Psi_n(x)e^{inx_1/R}$. Recall that our fermions are periodic 
and therefore, unlike a thermal Matsubara decomposition, the $n$ take 
values in the integers. 
In the background of the Wilson line \eqn{truevac}, the kinetic term 
for the fermion is
\be
\Psi^\dagger\gamma_u\left(\partial_\mu-i(A_\mu^1-A_\mu^2)\right)\Psi 
=\sum_{n}\Psi^\dagger_{-n}\left(\gamma_i\partial_i-i\gamma_1\left(
\frac{n}{R}+m\right)
\right)\Psi_n
\label{massive}\ee
We see that the effect of the mass term $m=1/2R$ is 
to shift the moding of the fermions to half-integers. The 
gauge fields remain integer moded. Our theory has 
therefore driven itself to a thermal compactification. 
If we Wick rotate to  Euclidean signature, all correlation 
functions will be those of a high temperature equilibrium thermal field theory. 
The resulting physics is well known. We expect that, at small 
$R\ll \Lambda^{-1}$, 
no condensate forms and the discrete chiral symmetry is restored. In analogy 
with QCD, it seems likely that a phase transition occurs at $R=R_c$. Needless 
to say, it would be interesting to understand the nature of the phase 
transition. In QCD with $N_f$ massless flavours, it is known that the chiral phase 
transition is 
second order for $N_f=2$, and first order for $N_f\geq 3$ \cite{pw}. However, since 
this analysis is based on the breaking of the continuous $SU(N_f)_L\times SU(N_f)_R$ 
symmetry, we should be cautious in extrapolating to our theory with only a discrete 
chiral symmetry.

\subsubsection*{Fractional Instantons on ${\bf R}^3\times {\bf S}^1$ and the Condensate}

We have seen that, at least for $R\ll\Lambda$, the orbifold theory is dynamically 
driven to the vacuum \eqn{truevac} where the full non-abelian gauge symmetry is restored. 
In this vacuum, there are no semi-classical fractional instanton configurations and, 
indeed, we have argued above that the discrete chiral symmetry remains unbroken. 

What happens above the critical 
radius, $R>R_c$? Perhaps the Wilson line \eqn{myvac} or \eqn{myvac2} and the 
associated fractional 
instantons play an important role in the formation of the condensate? Interestingly, 
recent lattice simulations of $SU(2)$ gauge theory at finite temperature suggest that 
both the Wilson line and fractional instanton are indeed present at 
temperatures $T\leq T_c$ \cite{lat}. Motivated by this 
observation, in this section we examine the the fermionic zero mode structure of 
the available instanton solutions in the vacua \eqn{myvac} and \eqn{myvac2}, and 
calculate the corresponding condensate. Of course, this calculation is valid only 
in the regime $R\ll R_c$, and its value is correspondingly dubious. 

The fractional instantons that appear in theories compactified on 
${\bf R}^{2,1}\times{\bf S}^1$ are also known as ``calorons''. If the gauge group 
is broken to the maximal torus by a Wilson line, then each factor of $SU(N)_i$ plays host to 
$N$ ``minimal calorons'' \cite{leeyi} which carry only four bosonic zero 
modes (3 translation, and 1 gauge rotation). All other classical solutions are 
composed of these objects. 
Of these $N$ minimal calorons, $N-1$ are simply monopole solutions which are independent 
of $x^1$ and have action $4\pi(v_{a+1}-v_{a})/g^2$ for 
$a=1,\ldots,N-1$. These are accompanied by one further "Kaluza-Klein" monopole 
solution, related to the originals by an $x^1$-dependent gauge transformation. 
It has action $4\pi(2\pi-(v_N-v_1))/g^2$. For the Wilson lines \eqn{myvac} and 
\eqn{myvac2}, we see that each of these minimal calorons, in each gauge group, 
has action $8\pi^2/g^2N$.

The ability of a given instanton solution to contribute to 
$\langle\lambda\psi\rangle$ depends on the structure of its fermionic zero 
modes. In our case, the existence of these zero modes depends strongly on the 
choice of classical Wilson lines\footnote{At this point we differ from \cite{gs}. Note 
that, as described previously, we allow for large gauge transformations which are 
twisted under the ${\bf Z}_N$ center, which also differs from \cite{gs}. 
I thank Misha Shifman for discussions on these 
issues.}. To see this, let us examine the Callias index theorem 
for monopoles in the presence of fundamental fermions. 
The important point is the dependence of the zero mode on the Kaluza-Klein 
mass of the fermions, which was studied in the Appendix of \cite{berk}. 
This Kaluza-Klein mass is a term of the form $im\Psi^\dagger\gamma_1\Psi$ as
seen in \eqn{massive}  
which is not Lorentz invariant in $d=3+1$, but is permitted in $d=2+1$, where it is 
also known as a ``real mass''. 
The result of \cite{berk} is that each Weyl fermion in the fundamental representation 
of the gauge group donates a single zero mode to the $a^{\rm th}$ caloron if its 
real mass $m$ lies within the range $v_a< m < v_{a+1}$. 

Returning to the orbifold theory, consider a caloron in gauge group 
$U(N)_1$. It sees $N$ ``flavours'' of Weyl fermions transforming in the fundamental 
representation, with real masses determined by the Wilson line of $U(N)_2$. 
At this point, the utility of the pictorial representation of Figure 1 becomes 
apparent. A caloron solution in $U(N)_1$ is depicted by a arc stretching between two 
solid lines. Each dotted line corresponds to two Weyl fermions seen by this caloron. 
These Weyl fermions carry fermionic zero modes only if they intersect the caloron arc. 

What happens for fractional instantons in the untwisted vacuum \eqn{myvac2} of Figure 1B? 
As we can see from the picture, the dotted lines only 
barely touch the arc at the ends. From the gauge theory perspective, this corresponds 
to the limit in which the real mass coincides with the expectation value  
$m\rightarrow v_a$, at which point the number of fermionic zero modes jumps 
discontinuously. In fact, what happens is that the fermionic zero mode becomes 
non-normalizable at this point. One may suspect that this means that there 
is no caloron contribution to the fermion condensate in this vacuum. This 
suspicion is confirmed by an explicit computation of the condensate for 
arbitrary real mass, subsequently taking the limit as $m\rightarrow v^a$ (see 
the formulas in the second paper of \cite{dkmtv}).

What about the Wilson line \eqn{myvac} depicted in Figure 1A? The eigenvalues of the 
$SU(N)_1$ and $SU(N)_2$ gauge groups are interlaced around the circle. Thus each 
minimal caloron solution receives zero 
modes from only a single flavour of Weyl fermion: each carries two 
fermionic zero modes, one for $\lambda$ and one for $\psi$. 
In short, all minimal calorons are 
ideal candidates to give rise to the condensate \eqn{prediction}. 
Summing over all minimal caloron contributions, following \cite{swandur} closely, 
we find
\be
\langle\lambda\psi\rangle=2N\int {d^3X}d\Omega\,  
\frac{J_B}{(2\pi)^4}\int d^2\xi\,\frac{1}{J_F}
\,{\lambda}^{(0)}(X){\psi}^{(0)}(X)
\left(\frac{\Gamma}{\Gamma_0}\right)e^{-8\pi^2/g^2N+i\theta/N} 
\nn\ee
The overall factor of $2N$ reflects the number of minimal calorons in this model. 
Each has three position collective coordinates $X$ and a phase collective coordinate 
$\Omega$. The bosonic Jacobians for these were calculated in \cite{dkmtv,swandur} 
and give $J_B=2^6\pi^4R/N$. The two Grassmannian collective coordinates  
$\xi$ are saturated by the zero mode insertions of ${\lambda}^{(0)}$ and 
${\psi}^{(0)}$. The 
Jacobian for fundamental fermions was calculated in \cite{dkmtv}, but 
in fact partially cancels the integration $\int\,d^3Xd^2\xi$ over the zero modes, 
leaving behind a factor of $1/2\pi R$. Finally, we come to the determinants, $\Gamma$. 
In the background of a self-dual field configuration they may be written as 
\be
\Gamma=\left(
\frac{\det^{1/2}(-\Dslash_1\Dslash_1^\dagger)}{\det^\prime
(-\Dslash_1^\dagger\Dslash_1)}\right)\,\left(
\frac{\det^{1/2}(-\Dslash_2\Dslash_2^\dagger)}{\det^\prime
(-\Dslash_2^\dagger\Dslash_2)}\right)\, 
\det{}^{\prime\,1/2}\left(\begin{array}{cc} -\Dslash_f\Dslash_{f}^\dagger & 0 
\\ 0 & -\Dslash_f^\dagger\Dslash_f \end{array}\right)
\label{dets}\ee
The first two factors come from integrating over the ghosts (numerators) and 
gauge fields (denominators), with $D_i$ acting in the adjoint representation 
of the $U(N)_i$ gauge group. The final factor arises from integrating out 
the fermions, with $D_f$ in the bi-fundamental representation. All 
operators are evaluated on the background of the caloron and $\det^\prime$ 
denotes the removal of zero modes. $\Gamma_0$ is the same operator evaluated on 
the vacuum, was found in the previous section to be $\Gamma_0=1$ in the background 
\eqn{myvac}.


In the background 
of the caloron, $\Gamma$ is UV divergent. Introducing a cut-off mass scale $\mu_0$ 
through Pauli-Villars regularisation, the leading order contribution to the divergence 
arises from the truncation of the four bosonic and two fermionic zero modes. Since 
$4-\ft12\times  2=3$, we have $\Gamma=c\mu_0^3$, with some real coefficient $c$. Further 
corrections are suppressed by $1/R\mu_0$. Putting everything together, and taking the 
cut-off to infinity, we have
\be
\langle\lambda\psi\rangle=32\pi^2 c \mu_0^3e^{-8\pi/g^2N}e^{i\theta/N}
\label{answer}\ee
where $c$ is an $R$-independent constant which, for $c=1$, results in the 
claimed condensate \eqn{prediction}. The above condensate is also holomorphic 
in $\tau$. In the semi-classical regime, such holomorphy reflects the fact that 
certain correlators are saturated by pure instanton  
contributions. For supersymmetric theories, this is assured by the extra 
goldstinos that arise for non-BPS configurations. In the present case, holomorphy 
holds only in the dilute gas approximation, where a background of $n_+$ 
well-separated instantons and $n_-$ anti-instantons carries 
$(n_++n_-)$ (approximate) fermionic zero modes. This is too many to contribute to 
$\langle\lambda\psi\rangle$. 


\subsubsection*{Summary}

We have argued that super Yang-Mills and its orbifold exhibit qualitatively 
different behaviour when compactified on small circles of radius $R\ll\Lambda^{-1}$. 
Specifically, in super Yang-Mills the fermionic condensate is independent of 
$R$, while the orbifold theory undergoes a phase transition at some 
$R=R_c$, below which the discrete chiral symmetry is restored. 
This effect is due to the condensation of twisted operators, which take 
the orbifold theory away from the vacuum of its parent where the orbifold 
conjecture may be tested. Thus, the disappointing  
conclusion of this note is that compactification on a spatial circle is 
not a useful way to test the orbifold conjecture. 

A glimmer of hope remains however. We have shown that in the presence of the 
Wilson line \eqn{myvac}, the fractional instanton 
has the correct zero mode structure to give rise to the conjectured 
fermion condensate. Hints from the lattice \cite{lat} suggest that 
this Wilson line and the associated fractional instanton may play a role 
at $R>R_c$. 

It is perhaps worth making one final comment. For super Yang-Mills, the 
condensate is independent of $R$ for all $N$. Is there similar behaviour 
for the orbifold theory for $R>R_c$ at large $N$? 
Once again, we may look to the analogy with QCD, now at low temperatures. As the 
temperature is increased from zero, the chiral condensate is known to decrease 
as $T^2/f^2_\pi$ \cite{what}. Since $f_\pi\sim {\cal O}(N^{1/2})$, this suggests that, 
at low temperatures,  
the condensate does indeed remain constant in the large $N$ limit.

\subsubsection*{Acknowledgments}
I'd like to thank Luis Bettencourt, Jiunn-Wei Chen, John Negele, York Schroder, 
D.T. Son, Matt Strassler, Jan Troost and especially Ami Hanany and Andy Neitzke for many 
useful discussions. I'd also like to thank Allan Adams and Misha Shifman for 
extensive comments and discussions on the draft. I'm supported by a 
Pappalardo fellowship, and am very grateful to Neil Pappalardo for the money. 
This work was also supported in part 
by funds provided by the U.S. Department of Energy (D.O.E.) under 
cooperative research agreement \#DF-FC02-94ER40818.

\end{document}